\documentclass[showpacs,twocolumn,prl,linenumbers]{revtex4}%
\usepackage[utf8]{inputenc}
\usepackage[T1]{fontenc}
\usepackage{amsmath}
\usepackage{amssymb}
\usepackage[pdftex]{graphicx} 

\newcommand{\abs}[1]{\left| #1 \right|} 
 
\newcommand{\ket}[1]{\left| #1 \right>} 
\newcommand{\bra}[1]{\left< #1 \right|} 
\let\baraccent=\= 
\renewcommand{\=}[1]{\stackrel{#1}{=}} 

\begin{document}
\title{ Microwave signature of topological Andreev-level crossings in a bismuth-based Josephson junction.
}

\author{ A. Murani$^{1,2,\ddag}$, B. Dassonneville$^{1,*}$, A. Kasumov$^{1}$, J. Basset$^{1}$,  M. Ferrier$^{1}$,  R. Deblock $^{1}$, S. Gu\'eron$^{1}$ and H. Bouchiat$^{1}$ }

\affiliation{$^{1}$ Laboratoire de Physique des Solides, CNRS, Université Paris-Sud, Université Paris-Saclay, 91405 Orsay Cedex, France
$^{2}$ Quantronics Group, Service de Physique de l’État Condensé (CNRS UMR 3680), IRAMIS, CEA-Saclay, 91191 Gif-sur-Yvette, France
$^{*}$ Current address: National Institute of Standards and Technology, Boulder, Colorado, 80305, USA
}

\begin{abstract}
Demonstrating the topological protection of Andreev states  in Josephson junctions  is an experimental challenge. In particular the telltale $4\pi$ periodicity expected  for the current phase relation  has remained elusive, because of fast parity breaking processes.  It was predicted that  low temperature   ac susceptibility measurements 
could reveal the topological protection of  quantum Spin Hall  edge states \cite{Fu2009}, by probing their low energy  Andreev spectrum.
We have performed such a microwave probing of the Andreev spectrum of a  phase-biased Josephson junction built around a bismuth nanowire, which was previously shown to host one-dimensional ballistic edge states. We find absorption peaks at the Andreev level crossings, whose temperature and frequency dependences point to  protected topological  crossings  with an accuracy limited by the electronic temperature of our experiment. 
\end{abstract}
\maketitle


One of the striking properties of topological matter is the existence of protected metallic states at the interfaces between two insulators with different topological invariants. Those states have a unique  dispersion relation: they display crossings of spin-momentum-locked Kramers partners at high symmetry points of the Brillouin zones, whose protection stems from the high spin-orbit interaction (SOI). Topological protection consequently allows for 1D ballistic transport (see e.g. \cite{Hasan} for a review). When superconducting correlations are induced in a  topological insulator (TI), particle-hole symmetry and fermion parity conservation enforce protected crossings of the Andreev eigenenergies at zero energy, which is often discussed in terms of  Majorana  States \cite{Fu2009,  Alicea2012, ReviewAguado}, in contrast to avoided crossings of  Andreev levels in   topologically trivial materials. In this paper, we demonstrate 
 a protected crossing in a crystalline Bi nanowire connected to two S electrodes (a S-Bi-S junction) using a high frequency linear response experiment, confirming the second order topological character of bismuth \cite{neupert}.

Crystalline bismuth, despite its semi-metallic character, has been shown \cite{Schindler} 
to belong to the recently discovered family of  higher order topological insulators.  Second order Topological Insulators  are insulating both in the bulk  and at high symmetry surfaces, but possess metallic 1D channels at the hinges between  surfaces  with different topological indices \cite{neupert}. The hinge states are helical and ballistic just like edge states in 2D topological insulators (2DTI). 
 The recent  prediction that bismuth belongs to this class of second order topological materials   explains  previous scanning tunneling microscopy  experiments revealing 1D states along the edges of hexagonal pits in Bi (111) crystals    \cite{Yazdani2014}, as well as transport  experiments on  Bi nanowires \cite{Li2014, Murani2017a}   proximitised by superconducting contacts.  Indeed, because of    electron and hole pockets at bismuth's Fermi energy, the few hinge states are bond to coexist with many non-topological bulk  and  surface states. In contrast with the  ballistic hinge states, those non-topological states are sensitive to disorder, resulting in diffusive motion of the charge carriers.  There is therefore no visible signature of topological transport in  a  Bi nanowire connected to non superconducting contacts, since the conductance  is  dominated  by the contribution of the diffusive channels.  The situation is fundamentally  different when superconducting electrodes (S) connect the Bi nanowire. The supercurrent through the S/Bi/S junction then runs preferentially along the wire's narrow hinge states, as revealed by the magnetic field periodic interference pattern originating from the hinges' spatial separation \cite{Li2014, Murani2017a}, similar to  Superconductor/2DTI/Superconductor junctions \cite{Hart2014, Pribiag2015}.

We have recently demonstrated the ballisticity of the hinge states over distances above one micrometer via the measurement of a sawtooth-shaped current-phase relation (CPR) of a S/Bi/S junction\cite{Murani2017a}.
Those experiments could not however demonstrate the topological nature of these hinge states since the sawtooth CPR was slightly rounded and the $4\pi$ periodicity  expected of a protected crossing was not observed. 
In fact, it is by now well understood that the $4\pi$ periodicity, a hallmark of topological  Josephson junctions, cannot be observed in dc CPR measurements \cite{pikulin}. Two physical phenomena  restore the $ 2\pi $ periodicity in our experiment:
one is  due to quasiparticle poisoning  which induces transitions between states of different parities at a given edge \cite{pikulin}, the other is due to  the  coupling between  the hinge states  of same parity on opposite sample edges  \cite{Trauzettel}.
By contrast, signatures of  $4\pi$ periodicity were observed in ac Josephson effect measurements \cite{Rokhinson, Boquillon, Chuan}.  The interpretation of those  experiments is however delicate since non-adiabatic transitions in voltage-biased Josephson junctions \cite{setsupra} as well as  topologically trivial Andreev states  with energy close to zero\cite{Akhmerov} also lead to signatures of $4\pi$ periodicity.

\begin{figure}
\includegraphics[width=\linewidth]{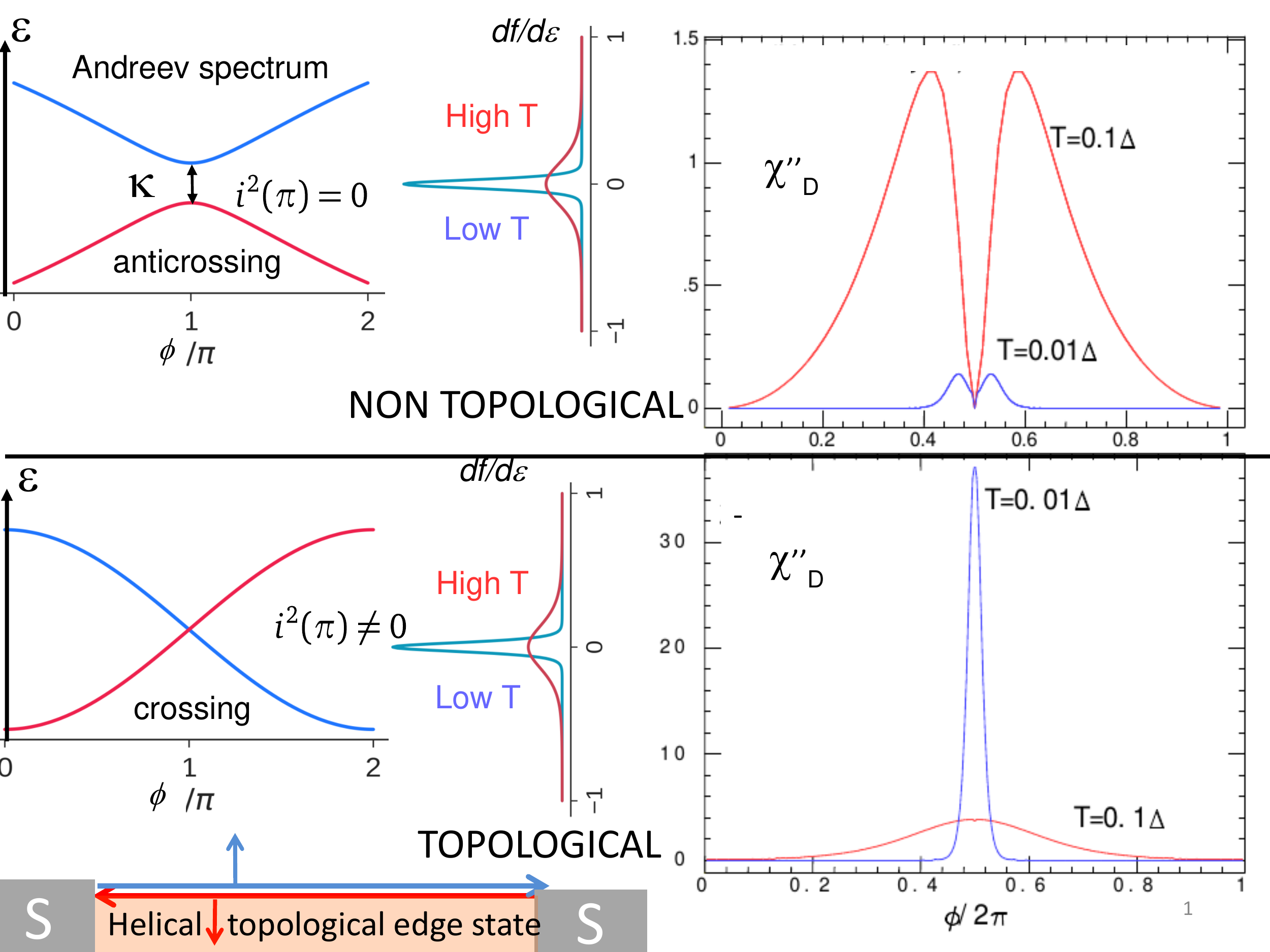}
\caption{Phase-dependent diagonal susceptibility as a signature of  Andreev level crossings.  Left: Sketch of a level crossing and anti-crossing at $\varphi=\pi$,  with the corresponding derivative of Fermi functions entering in the expression of $\chi_D(\varphi) \propto -i^2(\varphi) df/d\epsilon$.  Right: phase dependence of $ \chi''_D $  obtained  from the  tight binding computation of phase dependent  Andreev  bound states  in an SNS junction  on a hexagonal lattice  with on site disorder, in the  non-topological  (no SOI, upper panel) and  topological (next-nearest-neighbor SOI, bottom panel) regimes,  at  temperatures T=$0.01\Delta$ (blue)  and T=$0.1\Delta$ (red), with $ \Delta $ the superconducting gap (see ref. \cite{Murani2017b} for details). This contrast between a dissipation peak for the topological case and a minimum at $\pi$ for the non topological case is the basis of our experiment.
}
\label{fig1}
\end{figure}

An alternative proposal for the investigation  of topologically protected zero energy Andreev level crossings is to measure the ac  linear susceptibility of a phase-biased Josephson junction \cite{Murani2017b, Olesia}. In contrast to dc CPR measurements, 
ac susceptibility measurements  not only probe the Andreev spectrum (in particular level crossings) but in addition reveal the relaxation timescales of the spectrum occupation (diagonal density matrix elements) and interlevel transitions (off-diagonal elements)  \cite{Trivedi,dassonneville}. Specifically, the adiabatic, low frequency response is just the (non-dissipative) phase derivative of the CPR. At higher frequency, a non-adiabatic contribution to the susceptibility appears, $\chi_D$,  due to the relaxation of Andreev levels occupation. At low temperature, it is proportional to  
the highest occupied  Andreev level current $i$ and the phase derivative of its occupation    $\chi''_D(\varphi) \propto i \partial f /\partial\varphi= - i^2 \partial f / \partial \epsilon $ (where we have used the fact that the current carried by the Andreev level of energy $ \epsilon $ is $i=- \partial \epsilon /\partial\varphi$).
As a result,  a level crossing at zero energy translates into a  peaked dissipative response $\chi''_D$ at $\varphi=\pi$, which diverges at zero temperature.  
This result is  connected via the fluctuation-dissipation theorem to the prediction of  Fu and Kane  \cite{Fu2009} that the  phase-dependent  thermal noise of  the Josephson current in a topological junction  should peak at $\pi$. There is no such dissipation peak if the two levels anticross at $\pi$  (with a small gap $\kappa$), since then the current is zero,  and  both the noise and  ac dissipation are exponentially suppressed  at low temperature (below $\kappa$). This 
dichotomy demonstrates the power of high frequency linear susceptibility  and noise experiments to probe the  topological protection of edge or hinge states  in a  phase-biased topological insulator (see Fig.\ref{fig1} and Sup. Materials.). 

\begin{figure}
\includegraphics[width=\linewidth]{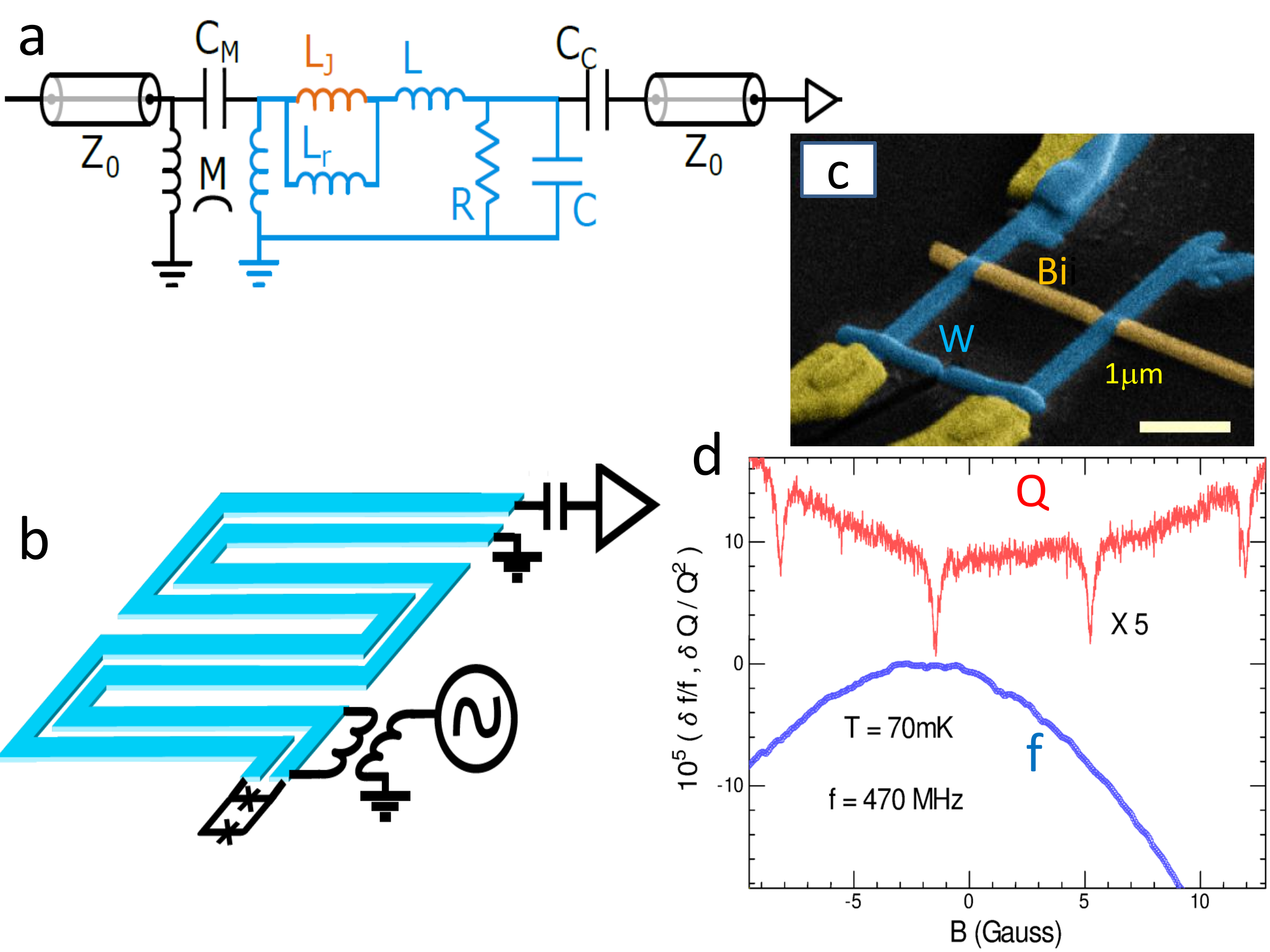}
\caption{(a)  and (b) Principle of the experiment: the Bi nanowire is modeled by an inductance $L_J$, the  W wire  including the constriction  is modeled with an inductance $ L_r$ in parallel with a resistance $R_r =1/G_{qp}$. The SQUID is inserted in the strip-line superconducting resonator measured in transmission with an inductive coupling to the microwave generator and a capacitive coupling to the cryogenic amplifier. 
(c)   Scanning electronic microscope  image of the Bi SQUID sample. 
(d)  Field induced variations of the quality factor  and frequency of the resonator's third eigenmode, at 70 mK  (average over 50 curves).  Note the sharp periodic absorption dips  on   Q(B) due to the Bi junction  whereas the smooth parabolic shift of f(B) is characteristic of the  field dependent penetration depth of the resonator's Nb meander lines.}
\label{fig2}
\end{figure}

 We have performed  such ac phase-biased  experiments by inserting an asymmetric SQUID built around a Bi nanowire into a multi-mode superconducting resonator (see Fig. \ref{fig2}).  We find periodic absorption peaks, whose temperature and frequency dependences point to topological crossings at $\pi$ of the Andreev levels,  to within  100 mK,  the  estimated electronic temperature of our experiment. This experiment also provides  the characteristic  relaxation time of Andreev levels occupation at $\pi$ caused by fermion  parity breaking due to quasiparticle poisoning.

The Bi nanowire-based asymmetric SQUID is connected to a $\lambda/4 $ multi-mode  resonator made of two parallel,  one meter long,  superconducting meander lines. The resonator is aligned to the asymmetric SQUID using standard e-beam lithography, followed by sputtering of $400 ~\rm{nm}$-thick Nb. We connect the resonator to the SQUID with focused-ion-beam-induced deposition of superconducting tungsten  (see Fig.2). The resonator is measured in transmission, in a dilution refrigerator with base temperature $50~\rm{mK}$, using homodyne detection. 
 The current's linear response $\delta I(t)=\delta I_\omega \exp -i\omega t$  to a  small time-dependent flux $\delta \Phi_\omega \exp -i\omega t$  is characterized by the complex susceptibility $\chi(\omega) =\delta I_{\omega}/\delta \Phi_\omega = i\omega Y(\omega)$, where Y is the admittance of the NS ring.
The  phase-dependences of the susceptibility's real and imaginary parts, $\chi'(\varphi)$ and $\chi''(\varphi)$, are related to the change of the n-th resonance's frequency $\delta f_n(\Phi)$ and inverse quality factor $\delta \left[1/Q_n\right](\Phi)$   induced  by the dc magnetic flux $\Phi$  via:
\begin{equation}
\chi'(\varphi) = -\frac{L_R}{L_{W}^2}\frac{\delta f_n(\Phi)}{2f_n},  \chi''(\varphi) = \frac{L_R}{L_{W}^2}\delta\left[\frac{ 1}{Q_n}\right](\Phi)
\end{equation}
where $ \varphi$, the superconducting phase difference is related to the flux  via $\varphi=-2\pi\Phi/\Phi_0$ with $ \Phi_0=h/2e $,  $L_{W}$, $L_R$ are the inductance of  the W loop (including the W constriction),  $\simeq 100 pH$ and the resonator, $L_R\simeq 1\mu  H$.
We have previously conducted similar experiments on long SNS junctions in which the normal part N is a topologically trivial diffusive Au wire \cite{dassonneville}. In those experiments, the susceptibility evolved from an adiabatic regime at low frequency, in which the susceptibility was exclusively non-dissipative, given by the  phase derivative of the Josephson current, to a dissipative regime at higher frequency, with minimal dissipation at $\pi$  in agreement with theoretical predictions \cite{Virtanen2011,Ferrier2013,Tikhonov2015}.
We report below a radically different behavior for the S/Bi/S junction: an exclusively dissipative susceptibility, peaked at $\pi$, that is compatible with topological ballistic Andreev states. 

We measure the linear response for resonator eigenfrequencies ranging from 0.28 to 6.7 GHz. The response is periodic, with a period of $7~G$, corresponding to one flux quantum through the SQUID loop, as expected from the dc flux biasing we impose. The variations with field of the resonance frequency and quality factor are  shown in Fig. 2d for the resonator's third eigenfrequency, $ f_3= 474~\rm{MHz}$.  The eigenfrequency shifts parabolically with field, as expected from the Nb resonator's kinetic inductance, but does not display periodic modulation. Thus  at these frequencies, the response of the Bi/S ring  is \textit{not} the flux derivative of the dc Josephson current previously measured by SQUID interferometry \cite{Murani2017a}. Such a contribution would be a detectable periodic modulation of $\delta f(\Phi)$, as demonstrated in Sup. Materials. 
In contrast, the quality factor displays below $0.5  ~\rm K$ and for all eigenfrequencies, clear periodic dips that correspond to dissipation peaks in $\chi''$ at  odd multiples of $ \Phi_0/2$ through the Bi-SQUID loop (i.e. a phase difference  equal to $\pi$). 

The height of the dissipation peak  $\delta_\pi(1/Q)$ at $\pi$  varies as $1/T$, with no observable saturation down to $100 ~\rm {mK}$ (see Fig. 3). It also increases linearly with frequency up to 4 GHz. Concomitantly,  the peak  width increases linearly with T and is independent  of frequency. Thus the  dissipation peak area is linear in frequency, with no temperature dependence.  We show below that those results are consistent with the expected dissipative linear response of a two level Andreev spectrum with a non-avoided crossing at zero energy and $\varphi=\pi$.
 
Indeed, such an Andreev spectrum has the form $\epsilon(\varphi)=\pm \epsilon_T(\varphi/\pi-1)$ near $\pi$, with $ \epsilon_T$ the Thouless energy, estimated to $ \epsilon_T\sim 4~{\rm K} $ from dc measurements. If we neglect the coupling between opposite edges of the wire, parity constraint and ballisticity  impose that there is no coupling by the current operator between the levels and therefore no allowed  interlevel transitions. The linear response's dissipative term $\chi''$  must thus be restricted to its diagonal term $\chi''_{\rm D}$, which is caused by the relaxation of  thermal occupations  of Andreev levels (see Sup. Materials). It reads $ \chi''_{\rm D} = -i_0^2\frac{\omega \gamma}{\omega ^2+ \gamma^2}(\partial f / \partial \epsilon) $ which, using the previous expression for the spectrum, yields:
\begin{equation}
   \chi''_{\rm D}= i_0^2\frac{\omega \gamma}{\omega ^2+ \gamma^2}\frac{1}{T\cosh^2\left(\frac{\epsilon_T}{2T}(\varphi /\pi-1)\right)}.
  \label{eqchi}
\end{equation}
 Here  $\gamma$ is the relaxation rate of the Andrev levels occupation and  $i_0 =\epsilon_T/\Phi_0$ is the  current carried by the Andreev states (in the long junction limit where $\epsilon_T$ is smaller than the superconducting gap). We note that this expression for the dissipative response is equivalent, via the fluctuation dissipation theorem, to the prediction of Fu and Kane for  the noise  power spectrum  $S(\omega)$ through  $ S(\omega) =4k_BT\chi''_D/\omega$. 
Fig. 3a shows how well the simple expression (\ref{eqchi}) fits  the experimental results, in particular the peaked  dissipation response at $\pi$ whose peak height and  inverse width are both proportional to 1/T down to 100 mK.  We show in  Sup. Materials that an avoided crossing at $\pi$ due to a small coupling $\kappa$ between levels at $\pi$, would generate (because of the current   going  to zero) a split peak around $\pi$   exponentially suppressed at temperatures below $\kappa$. Concomitantly, this coupling would also  allow interlevel transitions,  leading to an extra absorption peak  at $\pi$ whose width would be proportional to  $\kappa$ and  independent of temperature.  Since we see neither peak splitting nor temperature independent peak width, we conclude that there  is a perfect level crossing to within our experimental energy  resolution  of $100~mK$.
 
   We note that we have so far considered the contribution of only one pair of Andreev levels i.e. a single hinge state, whereas two hinges   carry the supercurrent (one at each acute angle)\cite{Murani2017a}.  Those two hinges must be coupled at  least at the  wire ends where they are both contacted to the superconductor. Using a distance between edges  $W_{Bi}\simeq 200$ nm and  a superconducting coherence length $\xi_W \simeq 20$ nm, we estimate this coupling to be  $\kappa=\epsilon_T \exp (-W_{Bi}/\xi_W) \simeq 0.2$ mK, which is about 500 times smaller than the base temperature of our dilution refrigerator.	This justifies our approximation of uncoupled hinge states. 
	In addition our  previous experiments \cite{Murani2017a}  indicate that one edge  carries a current 4 times larger than the other and therefore yields the main contribution to   $\chi_D$ by a factor 16.
	
       \begin{figure}
\includegraphics[width=\linewidth]{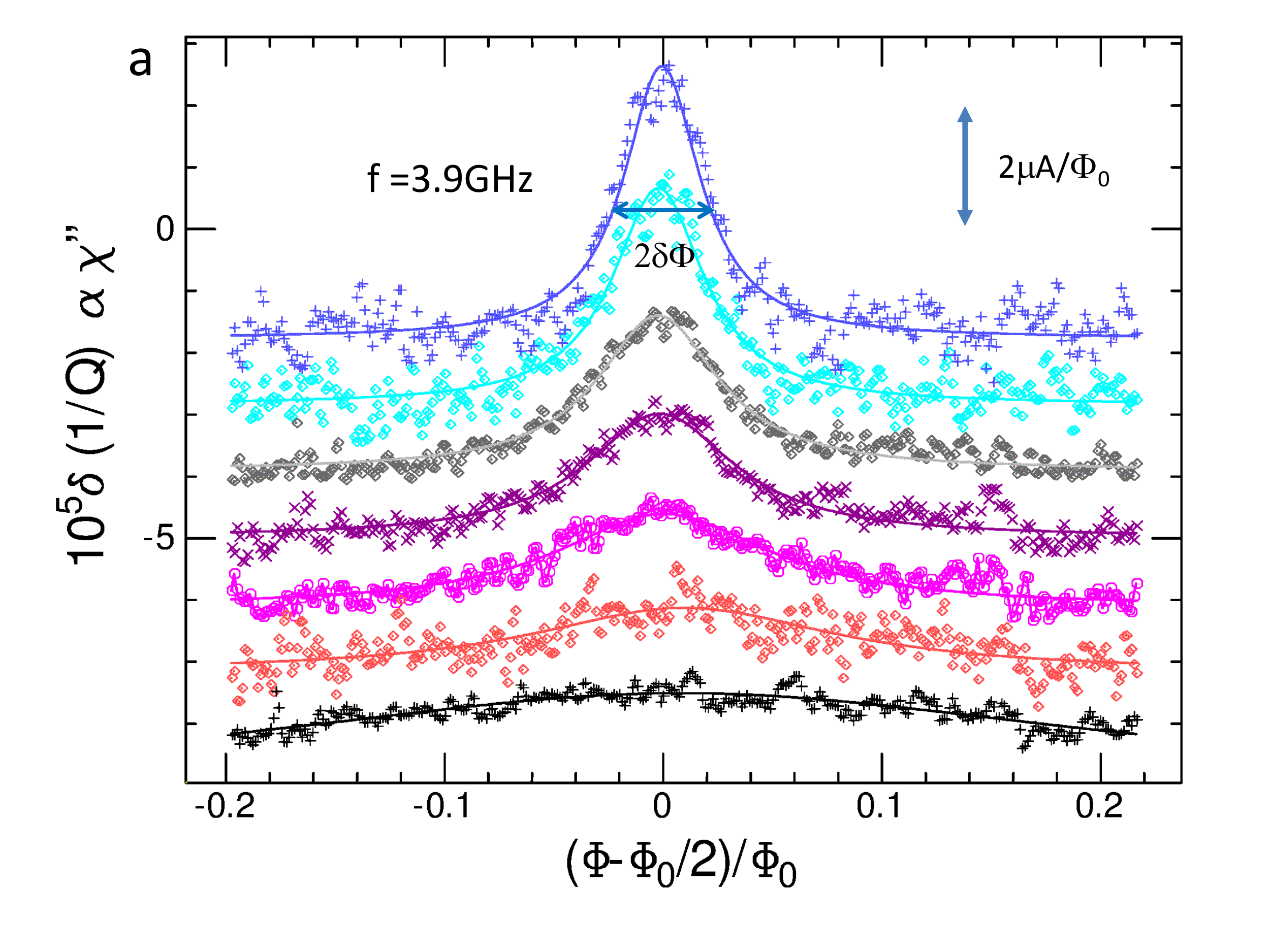}
\includegraphics[width=\linewidth]{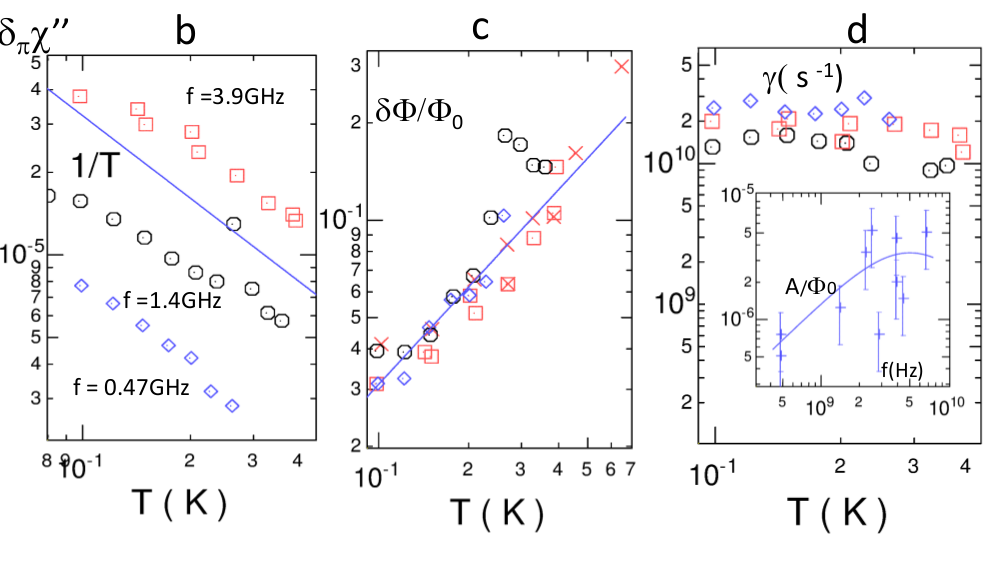}
\caption{(a) Temperature dependence  of the  dissipation peak around $\pi$ in $\chi''(\varphi)$  at f=3.9GHz and T=0.1, 0.15, 0.21, 0.27, 0.33, 0.39, 0.54 K  from upper to  lower curve (points are experimental data, solid line  are fits of eq. 2 to data) . (b) and (c)  Temperature dependence of   amplitude and width  $\delta_{\pi}(\chi'')\propto\delta_{\pi} 1/Q$ of  the dissipation peaks in $\chi''(\varphi)$,  at $\pi$, for different  eigenfrequencies of the  Nb resonator (diamonds 0.47 GHz, circles 1.4 GHz, squares 3.4 GHZ, crosses 4.5 GHz). 
 (d) Main panel: relaxation rate $\gamma $ deduced from the experiments at the three frequencies of panel b. Inset: frequency dependence of  the  absorption peak area measured at 100 mK. A reasonable fit with Eq. \ref{omegadep} is obtained taking $\gamma = 3$ $10^{10} s^{-1}$, despite the dispersion in the data,  mostly due to  uncertainties in  frequency-dependent calibration parameters.  }

\end{figure} 
 We have shown that the peaked  $\chi''_{\rm D}$ signals an unavoided level crossing at $\pi$ and a thermal occupation of the levels. We now discuss the rate at which the relaxation to  thermal equilibrium  occurs.  Since spin-orbit  coupling prevents direct transitions between spin-locked Andreev levels within one hinge, and since the coupling between the hinges is negligible, the most effective relaxation mechanisms must be due to quasiparticle poisoning by spin-degenerate unpaired quasiparticles. Such quasiparticles could originate either from the superconducting W \cite{Basset2018} or from non-topological (surface or bulk) states in the bismuth wire. We extract the relaxation rate from the frequency dependence of the  dissipation peak area  $ A (\omega)= \delta _\pi \chi'' \delta \Phi$,
\begin{equation}
  A (\omega)= i_0^2\omega\gamma/(\omega^2 +\gamma^2),
	\label{omegadep}
	\end{equation}
using  $i_0=400~\rm{nA}$ determined in the switching current experiment of \cite{Murani2017a}. This yields a relaxation rate $\gamma \simeq 2 \pm 1~10^{10} s^{-1}$, that is temperature-independent up to $0.6~\rm K$, see Fig. 3d.  Fu and Kane \cite{Fu2009}  suggested that  $\gamma(\pi, T)$ is the exchange rate 
between the zero energy Andreev states $\Psi_A (\pi)$  of the W/Bi/W junction and  quasiparticles at finite energy.  In a hard gap superconductor, this rate  is  exponentially suppressed at  temperatures below the  gap \cite{Yeyati,Marcus}. 
Our observation  that $\gamma$  is independent of temperature  below 0.5K indicates  the presence of quasiparticles at low energy in the  circuit. 
Following \cite{Yeyati},  $ \gamma $ can be deduced   from the Fermi golden rule: 
\begin{widetext}
	\begin{equation}
	\gamma=2 \pi^2\int  n_{qp} ( \epsilon)  \left(1-f(\frac{\epsilon}{k_BT_{el}})\right)  f_{BE}(\frac{\epsilon}{k_BT_{env}}) \abs{\bra{\Psi_A} {\bf I} \ket{\phi_{qp}(\epsilon)}}^2 \Re \left[Z(\frac{\epsilon}{\hbar})\right]\frac{d\epsilon}{\epsilon}
	\label{eqpoison}
	\end{equation}
\end{widetext}
  Absorption of a quasiparticle at energy $\epsilon$  gives rise to a photon emission  at the same energy in the electromagnetic environment of  the Bi junction  with a probability $P(\epsilon)$ proportional to  $\Re \left[Z(\frac{\epsilon}{\hbar})\right]$,  the  real part of the impedance   in parallel with the resonator,  $f$ and  $f_{BE}$ are the Fermi and the Bose-Einstein distribution functions, respectively taken at the electronic $ (T_{el})$ and environment $ (T_{env}) $ temperatures, $\bra{\Psi_A} {\bf I} \ket{\phi_{qp}(\epsilon)}$ is  the matrix element of the current operator  between the Andreev state and quasiparticle states.  
In Sup. Materials we estimate $\gamma$ from the quasi-particle conductance in parallel with  the kinetic inductance of the Bi wire, and the impedance of the resonator (coupled to the RF circuit). A value of $\gamma \simeq 10^{10} s ^{-1}$,  close to our experimental findings, is obtained if we take $T_{env}$ of the order of 2 K. This high effective temperature compared to the electronic temperature (100 mK) could be caused by the resonator's capacitive coupling to the cryogenic microwave amplifier (see Fig.2). 
We thus attribute the high relaxation rate $\gamma$ in our experiment to a sizable density of unpaired quasiparticles  in  the  SQUID and to the  dissipative component of the resonator  impedance at high frequency. These poisoning processes could in principle be considerably suppressed by using  a hard gap superconductor to contact the Bi nanowire and working with a single mode resonator with a narrow bandwidth \cite{Hays}.

There is one apparent inconsistency, however. The high relaxation rate means that $\omega/\gamma \leq 1$ for most eigen-frequencies we probe, and therefore the response regime should be quasi-adiabatic. This implies that $\chi'(\varphi)$  should be proportional to the derivative of the CPR that we measured on this very same sample in the previously reported experiment \cite{Murani2017a}. The fact that we detect no $\chi'(\varphi)$  may indicate that the even and odd parity levels are equally populated around $\pi$ because of the fast relaxation within one hinge. Since the two parity levels   carry opposite current, this would cancel $\chi'(\varphi)$ but not $\chi''(\varphi)$. In contrast, the CPR measurement experiment \cite{Murani2017a} was conducted at  low frequency ($10^4$ to $10^5$ Hz)  compared  to  the inter-hinge rate $\kappa/h\simeq 5$ $10^6 Hz $), so that during the  CPR measurement both edges can be explored, in practice lifting the helical feature, and restoring  the CPR of a long ballistic non-topological 1D wire \cite{Trauzettel,Fu2009}.



We have therefore obtained a consistent picture of the phase-dependent, high-frequency linear response of a Bi-based Josephson junction  whose  sharp dissipation peaks at $\pi$ reveal helical protected Andreev level crossings  and thus its topological nature. The short relaxation time we find (0.1 ns) is most likely due to subgap quasiparticle poisoning processes and to the coupling to an   insufficiently thermalized electromagnetic  environment.
The comparison between  dc and ac experiments suggests  a longer $\mu s$ inter-hinges scattering time. These results call for future measurements in the few  MHz range,  to explore fermion parity exchange processes between the opposite hinges \cite{Trauzettel}. Working instead at much higher frequency  (of the order of the Thouless energy) should enable to excite the parity conserving  transitions   in the long junction Andreev spectrum discussed in \cite {vanHeck2017, Glazman}
We acknowledge fruitful discussions with  M.Aprili, J. Aumentado, B. Doucot, H. Pothier, P. Simon and M. Triff on these experiments and their interpretation.

 \clearpage
    \onecolumngrid
\section{Supplemental Information}

\subsection{ Calculation of the susceptibility in the Kubo approximation }
We consider the situation where the phase dependent  Andreev spectrum is limited to two  time-reversed  quasi-ballistic Andreev states with a small anticrossing $\kappa$ at $\varphi=\pi$. The hamiltonian can be written via Pauli matrices $\hat \tau_i$, $1\leq i \leq3$.

 \begin{equation}
 {\cal \hat H} \equiv \epsilon_{\rm T}(\varphi-\pi)\hat \tau_3 + \kappa \hat \tau_1
 \end{equation}
 where $\epsilon_{\rm T}\equiv\hbar v_{\rm F}/L$ is the Thouless energy. 
 The current operator can therfore be written as:
 \begin{equation}
 {\cal \hat I} \equiv \frac{1}{\phi_0}\frac{\delta {\cal H}}{\delta \varphi} = i_0\hat \tau_3
 \end{equation}
 where $i_0 = \epsilon_{\rm T}/\phi_0$, $\phi_0 = h/2e$

 Diagonalization of $\cal \hat  H$ yields:
 \begin{equation}
 \begin{aligned}
 \epsilon_{\pm}(\varphi) &= \pm \sqrt{\epsilon_{\rm T}^2(\varphi-\pi)^2 + \kappa^2} \\
 |+\rangle &= \cos\left(\theta/2\right) |\uparrow \rangle + \sin\left(\theta/2\right) |\downarrow \rangle \\
 |-\rangle &= -\sin\left(\theta/2\right) |\uparrow \rangle + \cos\left(\theta/2\right) |\downarrow \rangle \\
 \end{aligned}
 \end{equation}
 The  matrix elements of the current operators in the $|+\rangle, |-\rangle$ basis read
 \begin{equation}
 {\cal \hat J} = 
 \left(
 \begin{matrix}
 i_0 \cos \theta & -i_0 \sin \theta \\
 -i_0 \sin \theta & -i_0 \cos \theta \\
 \end{matrix}
 \right)
 = \frac{i_0}{\sqrt{1 + (\frac{\kappa}{\epsilon_{\rm T}(\varphi-\pi)})^2}}
 \left(
 \begin{matrix}
 1 & -\frac{\kappa}{\epsilon_{\rm T}(\varphi-\pi)} \\
 -\frac{\kappa}{\epsilon_{\rm T}(\varphi-\pi)} & -1 \\
 \end{matrix}
 \right)
 \end{equation}
 In the following we set $\epsilon_{\rm T}=\hbar=k=1$ for simplicity.
The susceptibility of the NS loop is a function of matrix elements of the current operator between the eigenstates of the system:
 \begin{equation}
 \chi \equiv \frac{\delta I}{\delta \varphi_{\rm ac}} \quad\rightarrow \quad
 \left\lbrace
 \begin{aligned}
 \chi_{\rm D} &= -\frac{\partial f(\epsilon_-)}{\partial \epsilon}\frac{\hbar\omega}{\hbar \omega - i \gamma}\left|{\cal \hat J_{--}}\right|^2 \\
 \chi_{\rm N D} &= -\frac{f(\epsilon_-) - f(\epsilon_+)}{\epsilon_- -\epsilon_+}\frac{\hbar\omega}{\hbar \omega-(\epsilon_--\epsilon_+) - i \gamma}\left|{\cal \hat J_{-+}}\right|^2
 \end{aligned}
 \right.
 \end{equation}
The diagonal  and non diagonal contributions  $\chi_D$ and $\chi_{ND}$  describe respectively the relaxation of the occupations of the Andreev levels  and the interlevel transitions, and are  proportional to the diagonal and non-diagonal squared matrix elements of the current operator. 
  Using the e-h symmetry $\epsilon_- = - \epsilon_+ \equiv \epsilon(\varphi)$ we arrive at the following  formula:
 \begin{equation}
 \left\lbrace
 \begin{aligned}
 \chi_{\rm D} &= -i_0^2(\varphi-\pi)^2\frac{\omega}{\omega - i\gamma}\frac{1/T}{\epsilon(\varphi)^2\cosh^2\left(\epsilon(\varphi)/2T\right)}\\
 \chi_{\rm N D} &= -\frac{1}{2}i_0^2\kappa^2\frac{\omega}{\omega - 2 \epsilon(\varphi) - i\gamma}\frac{\tanh \left(\epsilon(\varphi)/2T\right)}{\epsilon(\varphi)^3}
 \end{aligned}
 \right.
 \end{equation}

 As seen in Fig.S1  these two contributions give rise to dissipative components $\chi''_D$ and $\chi''_{ND}$ with very  different  phase and temperature dependences. Whereas $\chi''_D(\phi)$ goes to zero at $\pi$ with a characteristic split \cite {lempitsky}, $\chi''_{ND}(\phi)$ exhibits a  peak at $\phi=\pi$ whose width is $ \kappa /\epsilon_{\rm T}$ and independent of temperature.  Assuming no temperature dependence of $\gamma$,  the temperature dependence of $\chi''_D$ is non monotonous and  determined by the  ratio $\kappa /T$ with a $1/T$ dependence for $T>\kappa $ and an exponential suppression at  low temperature $T<\kappa$. By contrast the T dependence of $\chi''_{ND}$   is determined by the Fermi functions centered at $\pm \omega$.  When $\omega \leq \gamma$ \cite{spectroscopy} the temperature dependence of is roughly constant  below  $\kappa$ and   varies like 1/T above.  We show in the main text that our experimental response can entirely to $\chi''_D$  and corresponds to a regime where  the level repulsion $\kappa$ is smaller than the  lowest electronic temperature  investigated, estimated to be 100mK. This illustrates the topological protection of hinge states of Bi .
\begin{figure}
\includegraphics[width=\textwidth]{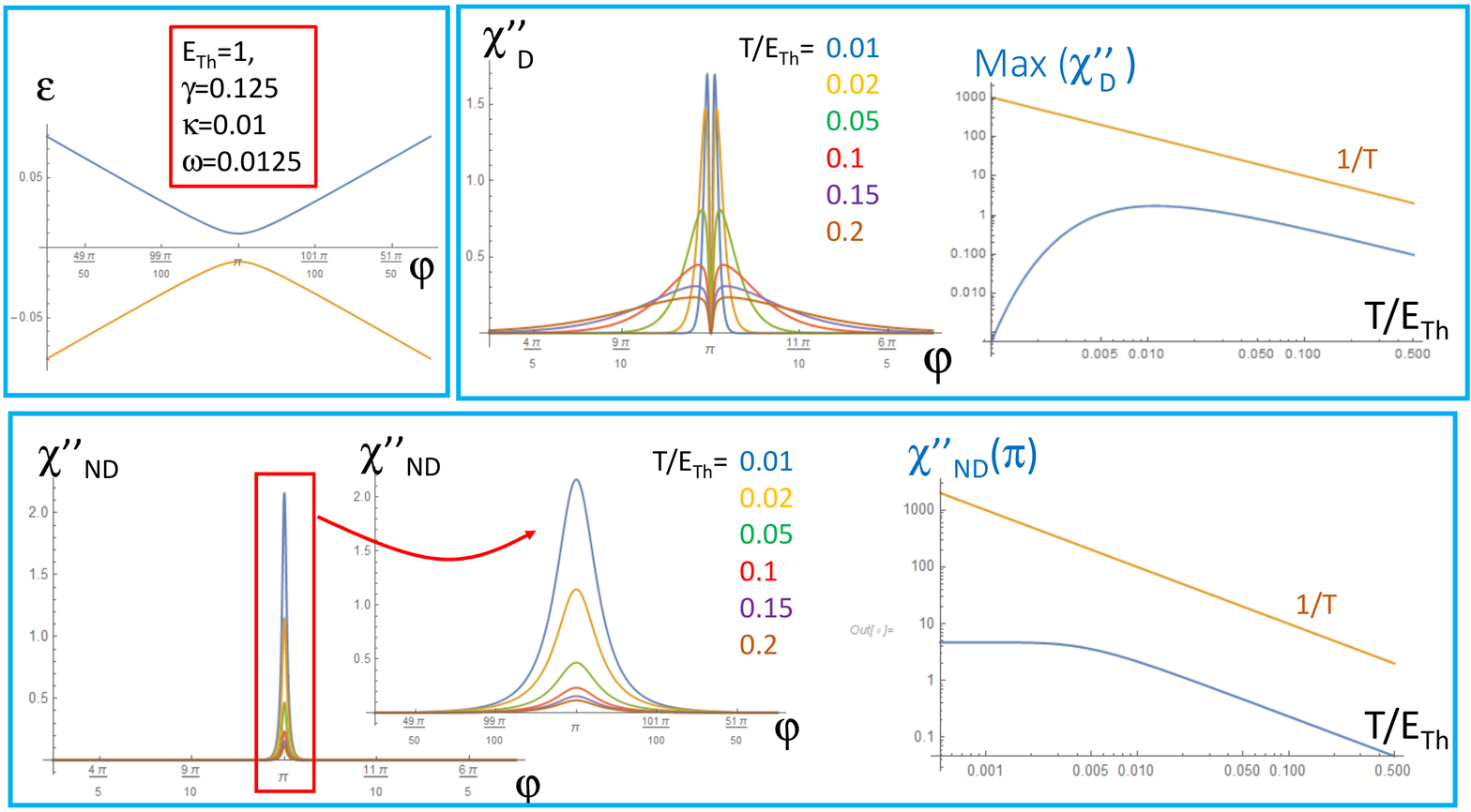}
\caption{ Phase  and temperature dependences of $\chi''_D$ (top curves)and $\chi''_{ND}$ (bottom curves) for two Andreev levels with a small coupling energy $\kappa$ . Note the characteristic   Lempitsky split peak in $\chi''_D$. We observe an exponential drop of  the maximum of$\chi''_D$ below $\delta$ whereas $\chi''_{ND}$ saturates at low temperature.
  }
\end{figure}

\begin{figure}
\includegraphics[width=\textwidth]{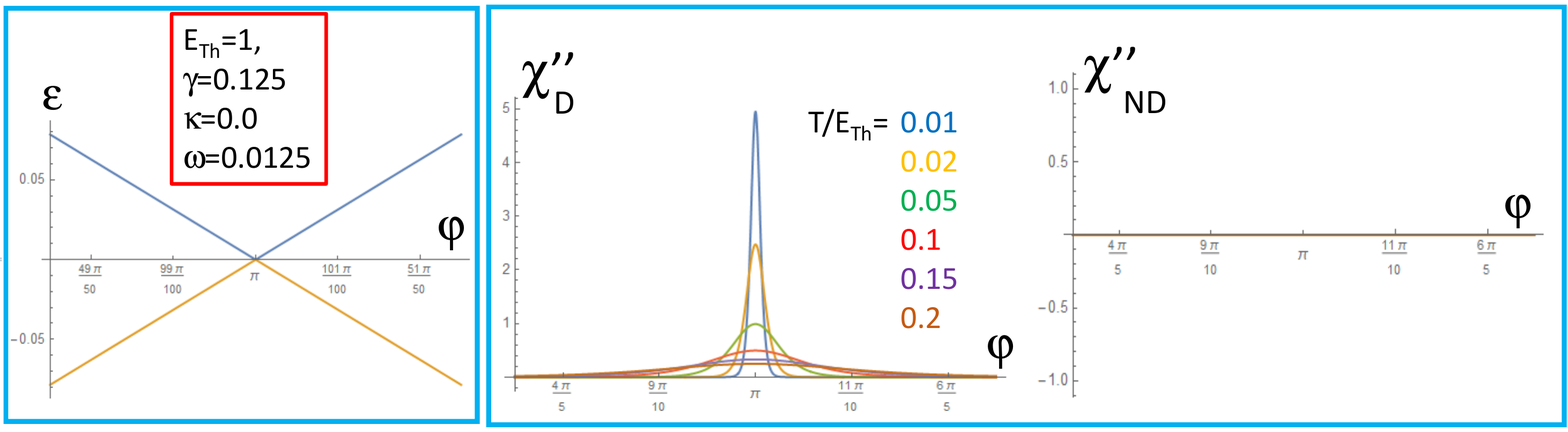}
\caption{ Phase dependence of $\chi''_D$ and $\chi''_{ND}$ in the limit of   perfect crossing between the Andreev levels ( $\kappa =0$). $\chi''_{ND}$ is equal to zero and 
 $\chi''_D$  exhibits a peak at $\pi$  which amplitude increasees as 1/T at low temperature and width  is proportionnal to T.
  }
\end{figure}

\subsection{Adiabatic response estimated from current phase relation measurements}

We have recently measured the phase-dependent switching current $I_S(\phi)$ of the presently investigated Bi SQUID before insertion in the superconducting resonator. The saw-tooth current-phase relation found in that  previous experiment is shown in Fig.S3 together with the expected  non dissipative response of the resonator in the adiabatic approximation:

\begin{equation}
 \left(\frac{-\delta f}{2f}\right) = \frac{L_{W}^2}{L_R}\chi'_{ad} (\phi)  
 \end{equation}

where $\chi'_{ad} (\phi)  =\partial I_S(\phi)/\partial \phi$, $L_{R}$ and $L_W$ are the inductances of the resonator and the W wire in parallel with the  Bi Josephson junction. This response would consist of periodic peaks on $\chi'_{ad}$ which are clearly not seen in the
experiment, see Fig.S3.  As we discuss in the main text of the paper this result is  at odd with the very short relaxation time of Andreev levels occupations we deduce from the analysis of the $\chi''(\phi,\omega)$ data. This is why we have to invoke a much longer relaxation time $\tau_c$  which is the time needed to couple the two nanowires edges destroying their helical character. It is reasonable to link this longer timescale   to  the very small coupling between the two edge statesthat run along opposite hinges  of the nanowire (see Fig. S3).  
\begin{equation}
1/\tau_c \simeq \epsilon_T \exp (-W_B/\xi_w)/\hbar\simeq 10^7 s^{-1}
\end{equation}
$\xi_w$ is the superconducting coherence length of  the superconducting  W  compound, which is of the order of 20nm, 10 times smaller than the  Bi wire  width $W_B$ which is 200nm.
This estimated value of $\tau_c$ can explain that the current phase relation measured at low frequency, below $10^7$ Hz is sensitive to the coupling between these 2 edge states  leading to the saw tooth current phase  relation  characteristic of a   ballistic 1D  long non-topological junction. By contrast the high frequency  response  probes a single helical edge yielding when poisoning relaxation is fast a parity averaged response $\chi'$ that is phase independent.

\begin{figure}
\includegraphics[width=\textwidth]{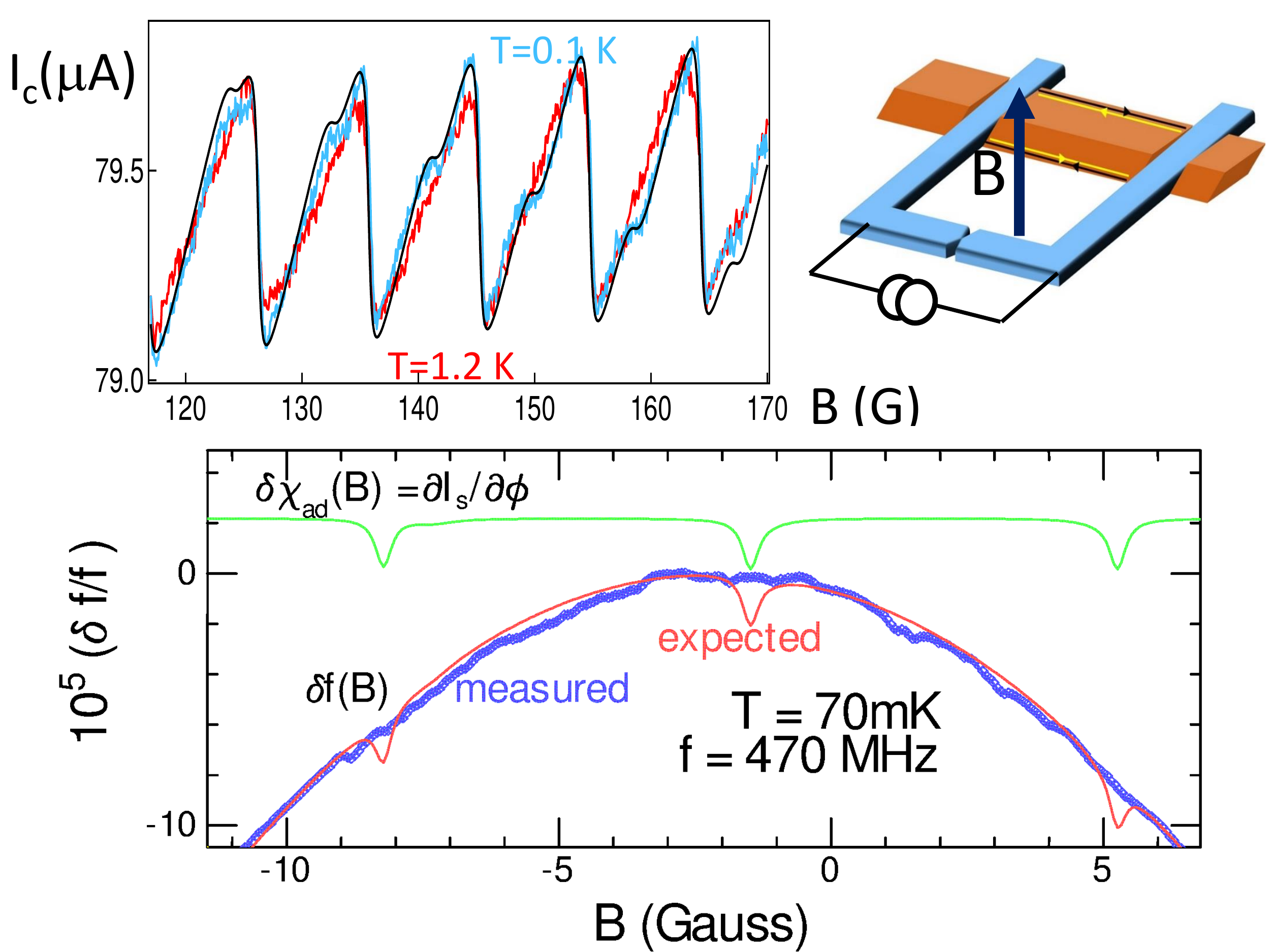}
\caption{ Green continuous line: expected adiabatic response of the in-phase susceptibility of the Bi SQUID.  Red continuous line: this signal si added to the bare response parabolic response of the reonator to  yield  the expected signal.  Comparison with  the  measured frequency shift   of the resonator's third resonance mode, at 70mK  (blue diamond points) demonstrates  that  the   expected periodic susceptibility peaks are not observed.  }
\end{figure}

\section{Estimation of the high frequency quasiparticle poisoning rate at $\pi$}
The aim of this section is  to give orders of magnitude  explaining the high dissipation rate of Andreev states rather than providing a rigorous calculation which is beyond the  scope of this work. 

 We start from eq.\ref{eqpoison} relating  from P(E) theory the   exchange rate  between the zero energy Andreev states and quasiparticles in the W/Bi/W junction to the impedance of the environment determined by the resonator \cite{Yeyati}.
\begin{equation}
\gamma= (2\pi/\hbar)\int  n_{qp} ( \epsilon)  (1-f(\epsilon/k_BT_{el})) f_{BE}(\epsilon/ k_BT_{env})\phi_0 ^2M_{qp} ({\rm Re}Z(\epsilon/\hbar) /\epsilon R_Q)d\epsilon
\label{eqpoison}
\end{equation}
with  $ M_{qp}=|< \Psi_A| {\bf I} |\phi_{qp} (\epsilon)> |^2$. This product  is simply related to   the quasiparticle  dimensionless conductance in parallel with the Bi SQUID through the Kubo formula $ n_{qp} M_{qp} (\epsilon) = g_{qp}(\epsilon)$. There are two possible origins for these quasiparticles which can come either from the W wire which has been shown  from microwave experiments to exhibit  a residual resistance of $10 k\Omega$ per micron \cite{Basset2018}, or from the  much smaller resistance of surface states of the Bi nanowire which do not carry
 the supercurrent. The coupling between  these surface states and protected edge states  is a priori  small, which makes it difficult to estimate their contribution.
We therefore assume that the combined contribution of the Bi and W wires to $g_{qp}$ at low energy is of the order of few units. The next step is to estimate the impedance of the environment Z. From fig.2, we can modelize Z by the impedance of the resonator $Z_R$ in parallel with the inductance of the W wire  in series with the  small coupling capacitance $\Gamma_c\simeq 10^{-13}F $ and the $Z_0= 50\Omega $ input impedance of the cryogenic preamplifier. This coupling imposes a temperature $T_{env}$ which we expect to be larger than 0.5 K, the largest temperature explored in these experiments. As a result in the frequency range where $L_W\omega \ll  Z_R \ll 1/(\Gamma_c \omega$,  $Z(\omega)$ writes:

\begin{equation}
{\rm Re}Z(\omega)= (Z_0 + {\rm Re}Z_R(\omega))L_W \omega^2\gamma_c^2
\end{equation}

Cutting the integral   in eq.\ref{eqpoison} over $\epsilon$ at $K_BT_{env}$  and replacing ${\rm Re}Z_R(\omega)$ by its average over $\omega$, $<Z_R>$, yields:
$\gamma =  {2\pi/\hbar} g_{qp}\left[ (Z_0 + <Z_R>)/R_Q \right]  L_W^2\Gamma_c^2 (k_BT_{env}/\hbar)^4$

Taking $<Z_R>=Z_0=50\Omega$, $T_{env}=2 K$, and $g_{qp} = 2$ leads to $\gamma = 10^{10} s^{-1}$, of the order of the experimental value.


\begin{thebibliography}{99}

\bibitem{Fu2009}
Liang and C. L. Kane, Phys. Rev. B 79, 161408(R)(2009).
\bibitem{Hasan} M. Z. Hasan and C. L. Kane, Rev. Mod. Phys. 82, 3045 (2010).
\bibitem {Alicea2012} Alicea J., Rep. Prog. Phys., 75  076501 (2012).                                        
\bibitem{ReviewAguado} R. Aguado, Riv. Nuovo Cimento 40, 523-593 (2017).
\bibitem{neupert} W. A. Benalcazar, B. A. Bernevig, and T. L. Hughes, Science 357, 61 (2017); F. Schindler, A. M. Cook, M. G. Vergniory, Z. Wang, S. S. P. Parkin, B. A. Bernevig,  and T. Neupert, Science Advances 4, 346 (2018).
\bibitem{Schindler} F. Schindler, Z. Wang, M. G. Vergniory, A. M. Cook, A. Murani, S. Sengupta, A. Y. Kasumov, R. Deblock, S. Jeon, I. Drozdov, H. Bouchiat, S. Guéron, A. Yazdani, B. A. Bernevig,  and T. Neupert,  Nat. Physics 14, 918 (2018).
\bibitem{Yazdani2014} Ilya K. Drozdov, A. Alexandradinata, Sangjun Jeon, Stevan Nadj-Perge, Huiwen Ji, R. J. Cava, B. A. Bernevig, Ali Yazdani, Nat. Phys. 10, 663 (2014).
\bibitem{Li2014}  Chuan Li, A. Kasumov, Anil Murani, Shamashis Sengupta, F. Fortuna, K. Napolskii, D. Koshkodaev, G. Tsirlina, Y. Kasumov, I. Khodos, R. Deblock, M. Ferrier, S. Gu\'eron, and H. Bouchiat, Phys. Rev. B 90, 245427 (2014).
\bibitem{Murani2017a} Anil Murani, Alik Kasumov, Shamashis Sengupta, Yu.A. Kasumov, V.T.Volkov, I.I. Khodos, F. Brisset, Raphaelle Delagrange, Alexei Chepelianskii, Richard Deblock, Helene Bouchiat, and Sophie Gu\'eron, Nat. Comm. 8, 15941 (2017).
\bibitem{Hart2014} S. Hart, H. Ren, T. Wagner, P. Leubner, M. Malbauer, C. Br$ \ddot{\rm u}$ne, H. Buhmann, L. W. Molenkamp, and A. Yacoby, Nat. Phys. 10 643, (2014).
\bibitem{Pribiag2015} V. S. Pribiag, A. J. A. Beukman, F. Qu, M. C. Cassidy, C. Charpentier, W. Wegscheider, and L. P. Kouwenhoven, Nat. Nano., vol. 10, 593 (2015).
\bibitem{Wiedenmann2016}  Wiedenmann, E. Bocquillon, R. S. Deacon, S. Hartinger, O. Herrmann, T. M. Klapwijk, L. Maier, C. Ames, C. Br$ \ddot{\rm u}$ne, C.Gould, A. Oiwa, K. Ishibashi, S. Tarucha, H. B$ \ddot{\rm u}$hmann, and L. W. Molenkamp,  Nat. Comm. 7, 10303, (2016).
\bibitem{pikulin}Beenakker CWJ, Pikulin DI, Hyart T, Schomerus H, Dahlhaus JP.,  Phys. Rev. Lett. \textbf{110}, 1 (2013). 
\bibitem{Trauzettel} F Crépin, B Trauzettel, Phys. Rev. Lett. 112 (7), 077002 (2014).
\bibitem{Rokhinson} Leonid P. Rokhinson, Xinyu Liu and Jacek K. Furdyna, Nat. Phys. 8,  795 (2012).
\bibitem{Boquillon} R. S. Deacon, J. Wiedenmann, E. Bocquillon,, F. Domínguez, T. M. Klapwijk, P. Leubner, C. Br$ \ddot{\rm u}$ne, E. M. Hankiewicz, S. Tarucha,, K. Ishibashi, H. Buhmann, and L.W. Molenkamp,	Phys. Rev. X \textbf{7}, 021011 (2017).
\bibitem{Chuan} Chuan Li, Jorrit C. de Boer, Bob de Ronde, Shyama V. Ramankutty, Erik van Heumen, Yingkai Huang, Anne de Visser, Alexander A. Golubov, Mark S. Golden, Alexander Brinkman, arXiv:1707.03154 (2017).
\bibitem{Akhmerov} Vuik, A, Nijholt, B, Akhmerov, A R, Wimmer, M, arXiv1806.02801.
\bibitem{setsupra} P. Joyez, Philippe Lafarge, A. Filipe, Daniel Esteve, M.H. Devoret,  Phys. Rev. Lett. 72, 2458 (1994). P.-M. Billangeon, F. Pierre, H. Bouchiat, R. Deblock, Phys. Rev. Lett. \textbf{98}, 216802 (2007).
\bibitem{Murani2017b} Murani A., Chepelianskii A., Gu\'{e}ron, S. and Bouchiat H.,  Phys. Rev. B {\bf 96}, 165415 (2017).
\bibitem{Olesia} Mircea Trif, Olesia Dmytruk, Helene Bouchiat, Ram\'on Aguado, Pascal Simon, Phys. Rev. B \textbf{97}, 041415(R) (2017).
\bibitem{Trivedi} Trivedi, N, Browne, DA, Phys. Rev. B \textbf{38}, 9581 (1988).
\bibitem{dassonneville} B.~Dassonneville,  M.~Ferrier,  S.~Gu\'{e}ron,  and H.~Bouchiat,  Phys. Rev. Lett. \textbf{110},   217001 (2013);  B. Dassonneville, A. Murani, M. Ferrier, S. Guéron, and H. Bouchiat, Phys. Rev. B \textbf{97}, 184505 (2018).
\bibitem{Virtanen2011}P.~Virtanen,  F.~S.~Bergeret,  J.~Cuevas, and  T.~Heikkil\"a,  Phys.Rev. B \textbf{83}, 144514  (2011).
\bibitem{Tikhonov2015} K. S. Tikhonov  and M. V. Feigelman, Phys. Rev. B \textbf{91}, 054519 (2015).
\bibitem{Ferrier2013}
M. Ferrier, B. Dassonneville, S. Guéron, and H. Bouchiat, Physical Review B 88, 174505 (2013).
\bibitem{Yeyati} D. G. Olivares, A. Levy Yeyati, L. Bretheau, C. Girit, H. Pothier, and C. Urbina. Phys. Rev. B \textbf{89}, 104504 (2014).

\bibitem{Marcus} A. P. Higginbotham, S. M. Albrecht, G. Kirsanskas,
W. Chang, F. Kuemmeth, P. Krogstrup, T. S. Jespersen,J. Nygard, K. Flensberg, and C. M. Marcus, Nature Physics 11, 1017 (2015).
\bibitem{Glazman} Jukka I. V\"ayrynen, Gianluca Rastelli, Wolfgang Belzig, and Leonid I. Glazman, Phys. Rev. B \textbf{92}, 134508 (2015).
\bibitem{vanHeck2017} B. van Heck, J. I. V\"ayrynen, and L. I. Glazman, Phys. Rev. B \textbf{96}, 075404 (2017).
\bibitem{Basset2018} J.Basset et al, in preparation.
\bibitem{Hays} M. Hays, G. de Lange, K. Serniak, D. J. van Woerkom, D. Bouman, P. Krogstrup, J. Nyg\aa{}rd, A. Geresdi, M. H. Devoret,  Phys. Rev. Lett. \textbf{121}, 047001 (2018).

\bibitem{lempitsky} This characteristic phase dependence of $\chi_D$ was initially derived by   Lempitsky for diffusive SNS junctions   S.V.  Sov. Phys. JETP 58, 624 (1983) and only measured a long time after in \cite{Dassonneville}.
\bibitem{spectroscopy} We remark that the frequency regime we are exploring $\omega \leq  \gamma $ is not the  regime of spectroscopy  experiments of Andreev states in quantum point contacts see: C. Janvier, L. Tosi, L. Bretheau, C. Girit, M. Stern,
P. Bertet, P. Joyez, D. Vion, D. Esteve, M. F. Goffman, H. Pothier, and C. Urbina, Science 349, 1199 (2015).






\end{thebibliography}
\end{document}